\documentclass[a4paper,twoside,12pt]{article}
\input{obsjkt.sty}

\newcommand{\degr}{\ensuremath{^\circ}}                           

\newcommand{\corot}{\textit{CoRoT}}
\newcommand{\kepler}{\textit{Kepler}}

\usepackage{rotating}

\begin{document} 

\OBSheader{Limb darkening laws implemented in \textsc{jktebop}}{J.\ Southworth}{2023 April}

\OBStitle{Re-parameterisation of four limb darkening laws and their implementation into the \textsc{jktebop} code}

\OBSauth{John Southworth}

\OBSinstone{Astrophysics Group, Keele University, Staffordshire, ST5 5BG, UK}


\OBSabstract{Limb darkening (LD) is typically parameterised using a range of functional `laws' in models of the light curves of eclipsing binary and transiting planetary systems. The two-coefficient LD laws all suffer from a strong correlation between their coefficients, preventing a reliable determination of both coefficients from high-quality light curves. We use numerical simulations to propose re-parameterisations of the quadratic, logarithmic, square-root and cubic LD laws that show much weaker correlations, and implement them into the {\sc jktebop} code. We recommend that these re-parameterisations are used whenever both LD coefficients are fitted. Conversely, when fitting for only one coefficient, the standard laws should be used to avoid problems with fixing coefficients at poor values. We find that these choices have little effect on the other fitted parameters of a light curve model. We also recommend that the power-2 LD law should be used as default because it provides a good fit to theoretical predictions, and that the quadratic and linear laws should be avoided because they do not.}


\section*{Introduction}

Limb darkening (LD) is a universal phenomenon which modifies the brightness of stars across their disc. LD results in a wavelength-dependent decrease in brightness from the centre of the observed disc to the limb, and in a steeper drop-off closer to the limb compared to near the centre. It arises because sightlines which enter the surface of the star at an angle (``slant viewing geometry'') penetrate less deep into the atmosphere, see cooler plasma than a perpendicular sightline, and so perceive a lower flux.

LD was first noticed in our Sun by Luca Valerio in 1612 \cite{Kopal59book}, and was first measured by Pierre Bouguer in 1729 \cite{Bouguer1760}. It must be accounted for in any observing project which involves spatially resolving a star, specifically interferometry, eclipsing binaries (EBs) and transiting planetary systems (TEPs). All analysis methods that the author is aware of for EBs and TEPs include a treatment of LD in order to properly represent the characteristics of the object(s) being considered.

In this work we describe the implementation of multiple LD laws into the {\sc jktebop}\footnote{\texttt{http://www.astro.keele.ac.uk/jkt/codes/jktebop.html}} code \cite{Me++04mn2,Me13aa} for modelling the light and radial velocity curves of EBs and TEPs. The novelty of this work lies primarily in the re-parameterisation of the two-coefficient LD laws to mitigate the strong correlations between the two coefficients. We begin with a reminder of the different LD laws in use, present the re-parameterisations we adopt, and conclude with advice on using the LD functionality now included in {\sc jktebop}.


\section*{Limb darkening laws}

For the analysis of the light curves of EBs, LD was implemented in the pioneering Russell-Merrill method \cite{Russell12apj,Russell12apj2,RussellShapley12apj,RussellShapley12apj2,RussellMerrill52book} using the linear law \cite{Russell12apj,Schwarzschild06}:
\begin{equation}
\frac{F(\mu)}{F(1)} = 1 - u_{\rm lin}(1-\mu) \,\,,
\end{equation}
where $F(\mu)$ is the flux at position $\mu = \cos\gamma$ on the stellar disc, $\gamma$ is the angle between the observer's line of sight and the surface normal, $F(1)$ is the flux at the centre of the disc, and $u_{\rm lin}$ is the linear LD coefficient. The strength of the LD is specified by $u_{\rm lin}$, which is normally between unity (no limb darkening) and zero (surface flux decreases to zero at the limb).

The linear LD law has been known for over a century to be an inadequate representation of the solar LD \cite{Very02apj,SchwarzschildVilliger06apj,PierceSlaughter77soph,NeckelLabs84soph} prompting more sophisticated laws to be proposed: the quadratic LD law (Kopal \cite{Kopal50}):
\begin{equation}
\frac{F(\mu)}{F(1)} = 1 - u_{\rm quad}(1-\mu) - v_{\rm quad}(1-\mu)^2 ~~,
\end{equation}
the logarithmic law (Klinglesmith \& Sobieski \cite{KlinglesmithSobieski70aj}):
\begin{equation}
\frac{F(\mu)}{F(1)} = 1 - u_{\rm log}(1-\mu) - v_{\rm log}\mu\ln\mu ~~,
\end{equation}
the square-root law (D{\'{\i}}az-Cordov{\'e}s \& Gim{\'e}nez \cite{DiazGimenez92aa}):
\begin{equation}
\frac{F(\mu)}{F(1)} = 1 - u_{\rm sqrt}(1-\mu)-v_{\rm sqrt}(1-\sqrt{\mu}) ~~,
\end{equation}
the cubic law (van't Veer \cite{Vantveer60book}):
\begin{equation}
\frac{F(\mu)}{F(1)} = 1 - u_{\rm cub}(1-\mu) - v_{\rm cub}(1-\mu)^3 ~~,
\end{equation}
the power-2 law (Hestroffer \cite{Hestroffer97aa}):
\begin{equation}
\frac{F(\mu)}{F(1)} = 1 - c(1-\mu^\alpha) ~~,
\end{equation}
and the four-parameter law proposed by Claret \cite{Claret00aa}:
\begin{equation}
\frac{F(\mu)}{F(1)} = 1 - \sum_{n=1}^4 u_n(1-\mu^{n/2}) ~~.
\end{equation}

The {\sc ebop} code \cite{Etzel75,Etzel81conf}, on which {\sc jktebop} is based, used the linear LD law. D{\'{\i}}az-Cordov{\'e}s \& Gim\'enez \cite{DiazGimenez92aa} and Gim\'enez \& D{\'{\i}}az-Cordov{\'e}s \cite{GimenezDiaz93conf} modified {\sc ebop} to include the quadratic and square-root LD laws. The current author subsequently added these and the logarithmic, cubic and four-parameter laws into {\sc jktebop} (versions 12, 15 and 31). We have now added the power-2 law ({\sc jktebop} version 43) which means that all the laws given above are now implemented in {\sc jktebop}. The cubic law was included specifically because it was expected that the greater functional difference between the two terms (compared to the quadratic law) would make the two coefficients less correlated; it is shown below that this is indeed the case. It is possible within {\sc jktebop} to use different LD laws for the two stars, with the exception of the four-parameter law.


\section*{Review of published re-parameterisations of the LD laws}

Our experience of using the LD laws in {\sc jktebop} for a wide range of EBs and TEPs is that: the linear law is adequate for most ground-based data but not for light curves from space missions such as \kepler, \corot\ and TESS; results from the two-parameter laws are typically in excellent agreement; one should fit for one of the two LD coefficients when possible because theoretical predictions are imperfect; fitting for both LD coefficients in the two-coefficient laws is not recommended because they can be severely correlated. Strong correlations are a particular issue for Markov chain Monte Carlo (MCMC) codes as they cause a long autocorrelation length and thus decrease the number of independent samples in the Markov chains. Support for these statements can be found in correlation plots \cite{Me++07aa,Me08mn} and supplementary material for the {\it Homgeneous Studies} publications \cite{Me08mn,Me10mn,Me11mn,Me12mn}. The strong correlations have also been noticed by other researchers, e.g.\ refs. \cite{Carter+08apj} and \cite{Pal08mn}.

The correlations could be decreased by changing the parameterisation of the LD laws, and a range of re-parameterisations have been proposed for the quadratic law. Brown et al.\ \cite{Brown+01apj} fitted for the sum and difference of the LD coefficients:
\begin{equation}
u^\prime = u_{\rm quad} + v_{\rm quad}
\end{equation}
\begin{equation}
v^\prime = u_{\rm quad} - v_{\rm quad} 
\end{equation}
Holman et al.\ \cite{Holman+06apj} used another:
\begin{equation}
u^\prime = 2\,u_{\rm quad} + v_{\rm quad}
\end{equation}
\begin{equation}
v^\prime = u_{\rm quad} - 2\,v_{\rm quad} 
\end{equation}
and P\'al \cite{Pal08mn} generalised these to
\begin{equation}
u^\prime = u_{\rm quad}\sin\theta - v_{\rm quad}\cos\theta
\end{equation}
\begin{equation}
v^\prime = u_{\rm quad}\sin\theta + v_{\rm quad}\cos\theta 
\end{equation}
where $\theta$ depends on the properties of the system being studied but is usually between 35\degr\ and 40\degr. Kipping \cite{Kipping13mn} has explored these in detail, and Howarth \cite{Howarth11mn} has discussed the comparison between observed and theoretical LD coefficients.

Maxted \cite{Maxted18aa} proposed a re-parameterisation of the power-2 LD law to depend on the coefficients $h_1$ and $h_2$ where
\begin{equation}
h_1 = \frac{F(0.5)}{F(1)} = 1 - c\,(1-2^{-\alpha})
\end{equation}
and
\begin{equation}
h_2 = \frac{F(0.5)-F(0)}{F(1)} = c\,2^{-\alpha}
\end{equation}
and $h_1$ and $h_2$ are only weakly correlated (see also Short et al.\ \cite{Short+19rnaas}).

We are not aware of proposed re-parameterisations for any of the other laws, a point also noted by Czismadia \cite{Csizmadia20mn}.


\section*{Data for numerical experiments}

It is desirable to avoid strong correlations between parameters when fitting the light curves of EBs and TEPs. We therefore chose to re-parameterise the two-parameter LD laws with coefficients that are less strongly correlated. As multiple differing options have been published for the quadratic law, and none for any of the other laws (except power-2), we decided to determine our own. The most straightforward way to do this is via numerical experiments.

We identified a set of five EBs and TEPs with a variety of properties and for which excellent light curves exist. The rationale for these choices is that we expected the correlations between LD coefficients to depend on the physical attributes of a given system so needed to include objects with a range of characteristics, and that very high-quality photometry is needed to fit for both LD coefficients in a given system.

The first object we analysed was the EB IT~Cas, which was chosen because it shows deep V-shaped eclipses which arise from two very similar stars with an orbital inclination near 90\degr, and thus should sample the full range of $\mu$ values on the stellar discs. For this we used the Simple Aperture Photometry (SAP) from sector 17 of the Transiting Exoplanet Survey Satellite (TESS) downloaded from the Mikulski Archive for Space Telescopes (MAST\footnote{\texttt{https://mast.stsci.edu/portal/Mashup/Clients/Mast/Portal.html}}). We used only data with a \texttt{QUALITY} flag of zero, ignored the data errors as they were too small, and rejected all data more than one eclipse duration from the midpoint of an eclipse in order to save computing time. A detailed analysis of this system is in preparation and will be presented in due course as part of the \textit{Rediscussion of Eclipsing Binaries} project \cite{Me20obs}.

\begin{figure}[t] \centering \includegraphics[width=\textwidth]{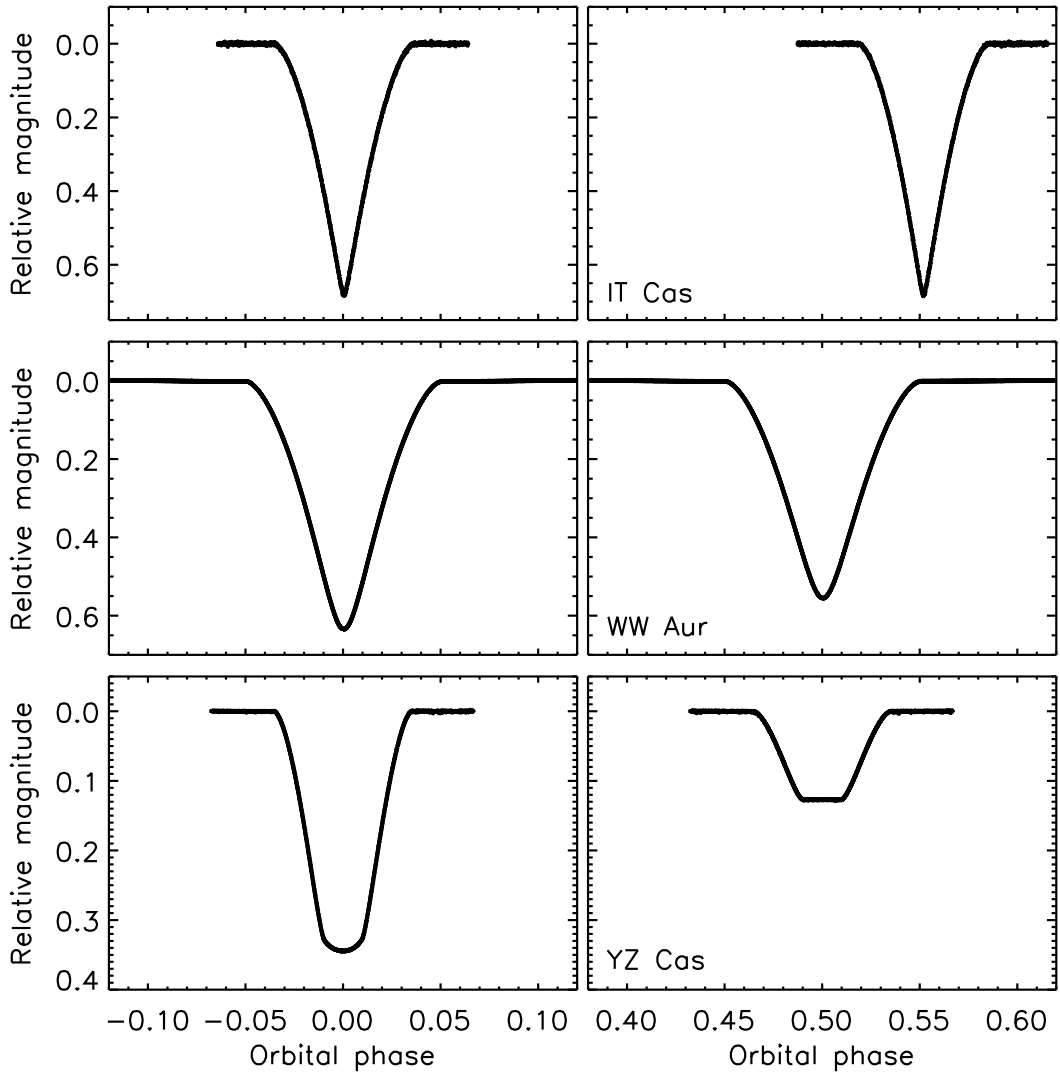} \\
\caption{\label{fig:lceb} TESS\ short-cadence SAP photometry of the three EBs analysed
in the current work. The primary eclipses are shown in the left panels and the secondary
eclipses in the right panels. The names are labelled on the panels.} \end{figure}

\begin{figure}[t] \centering \includegraphics[width=\textwidth]{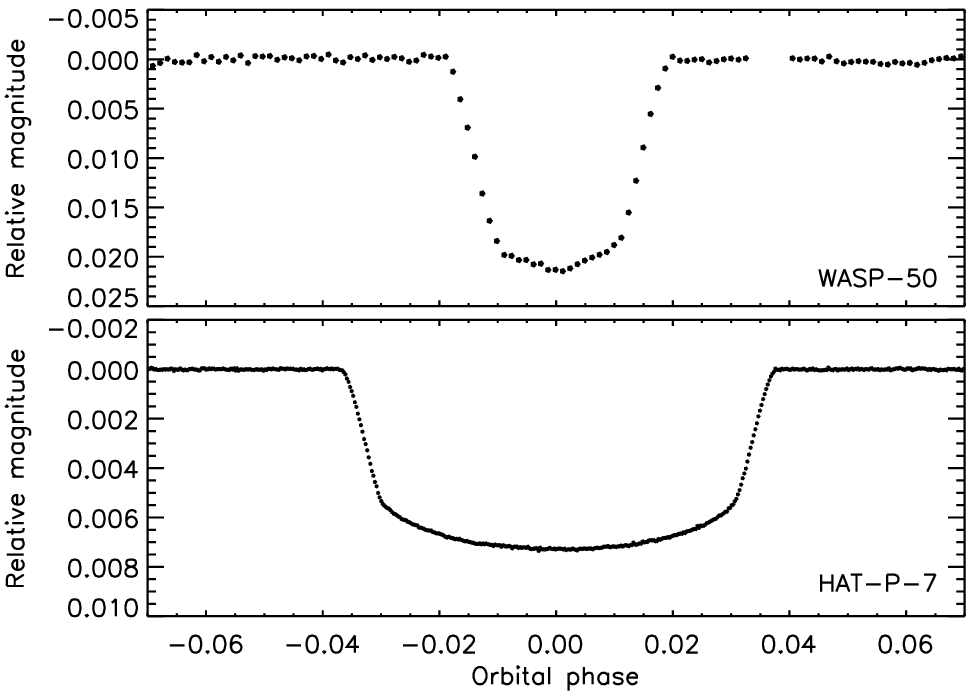} \\
\caption{\label{fig:lctep} Light curves of the two TEPs analysed in the current work
from the New Technology Telescope (WASP-50) and \kepler\ (HAT-P-7). } \end{figure}

For our second object we chose WASP-50, a TEP for which an extremely high-quality transit light curve is available from a ground-based telescope \cite{TregloanMe13mn}. These data proved to be useful but of insufficient quality to reliably measure two LD coefficients. We therefore chose a third object, the TEP system HAT-P-7 \cite{Winn+09apj3} for which an extraordinarily good light curve is available from the \kepler\ satellite \cite{Borucki16rpph}. We used the same data as in Southworth \cite{Me11mn}, which comprise the first 59 transits observed, all in short cadence mode \cite{Jenkins+10apj2}.

We also added a fourth object, the totally-eclipsing binary YZ~Cas \cite{Pavlovski+14mn}, for which we used the sector 19 data from TESS. Finally, after inspection of the preliminary results, we added WW~Aur \cite{Me+05mn} as it shows deep V-shaped eclipses similar to those of IT~Cas but has a circular orbit. For WW~Aur we used the sector 45 data from TESS. For both objects the TESS data were prepared in the same way as for IT~Cas. The light curves of the five objects are shown in Figs. \ref{fig:lceb} and \ref{fig:lctep}. We would have liked to extend this to stars hotter than YZ~Cas~A but were unable to identify a suitable candidate: all options we explored had either shallow eclipses, large fractional radii, pulsations, or no high-quality light curve.

The light curve of each object was modelled using {\sc jktebop} and a two-parameter LD law, with both LD coefficients fitted. Once a good fit was obtained, we ran a set of 1000 Monte Carlo simulations \cite{Me++04mn,Me08mn}, which comprised the generation and then least-squares fit of 1000 synthetic datasets with the same timestamps as the original data and brightness measurements taken from the original best-fitting model with Gaussian noise applied. This was performed for the quadratic, logarithmic, square-root and cubic LD laws. We did not consider the linear LD law, because it only has one coefficient so is not affected by correlations between coefficients, or the power-2 law, as the $h_1$ and $h_2$ approach was judged to be already satisfactory. Conversely, the four-parameter law exhibits such strong correlations between its coefficients that we considered it to be a lost cause so made no attempt to re-parameterise it.


\begin{table} \centering
\caption{\em \label{tab:corr} Linear Pearson correlation coefficients between the $u$ and $v$
coefficients of the two-parameter LD laws, assessed using Monte Carlo simulations as implemented
in {\sc jktebop}, for each of the five objects included in the numerical experimentation.}
\begin{tabular}{lcccc}
{\em Object}      & {\em quadratic law} &  {\em logarithmic law} & {\em square-root law} & {\em cubic law} \\[3pt]
IT~Cas            &  $-$0.982  &  $+$0.998  &  $-$0.999  &  $-$0.952   \\
WASP-50           &  $-$0.951  &  $+$0.994  &  $-$0.996  &  $-$0.605   \\
HAT-P-7           &  $-$0.992  &  $+$0.992  &  $-$0.999  &  $-$0.973   \\
YZ~Cas            &  $-$0.978  &  $+$0.987  &  $-$0.999  &  $-$0.914   \\
WW~Aur            &  $-$0.995  &  $+$0.997  &  $-$0.999  &  $-$0.985   \\[3pt]
\end{tabular}
\vskip10pt
\end{table}

\section*{New re-parameterisations}

We first assessed the linear Pearson correlation between the two LD coefficients in each Monte Carlo simulation, using the \texttt{correlate} function in IDL\footnote{{\tt http://www.harrisgeospatial.com/SoftwareTechnology/IDL.aspx}}. The results are given in Table~\ref{tab:corr} and support several conclusions. First, the correlations between $u$ and $v$ are in general horrendous. Second, we notice that the correlations are at their worst when the data are of the highest quality. Third, the coefficients of the square-root law exhibit almost perfect correlations so should never be fitted together. Fourth, the coefficients of the cubic LD law have the lowest correlations, supporting the expectation mentioned above.

\begin{table} \centering
\caption{\em \label{tab:x} Values of $x$ which minimise the correlation
between $u^\prime$ and $v^\prime$, for each LD law and each object studied.}
\begin{tabular}{lcccc}
{\em Object} & {\em quadratic law} & {\em logarithmic law} & {\em square-root law} & {\em cubic law} \\[3pt]
IT~Cas            &     0.44   &    0.75    &    0.57    &     0.19    \\
WASP-50           &     0.59   &    0.57    &    0.51    &     0.29    \\
HAT-P-7           &     0.62   &    0.60    &    0.62    &     0.39    \\
YZ~Cas            &     0.63   &    0.64    &    0.60    &     0.33    \\
WW~Aur            &     0.58   &    0.62    &    0.62    &     0.35    \\[3pt]
Adopted value     &     0.6    &    0.6     &    0.6     &     0.3     \\[3pt]
\end{tabular}
\vskip10pt
\end{table}

We next sought alternative parameterisations that would reduce these correlations. We chose a functional form that is similar to that of P\'al \cite{Pal08mn} but simpler:
\begin{equation}
u^\prime = u + x\,v
\end{equation}
\begin{equation}
v^\prime = u - x\,v
\end{equation}
where the quantity $x$ can be chosen to minimise the correlation between $u^\prime$ and $v^\prime$ for each LD law. The implementation of this in {\sc jktebop} was done by modifying the input and output sections but converting the LD to the original parameterisations when calculating a model datapoint. This meant that we needed only the inverse transforms, which can easily be shown to be:
\begin{equation}
u = \frac{u^\prime+v^\prime}{2}
\end{equation}
\begin{equation}
v = \frac{u^\prime-v^\prime}{2x}
\end{equation}
independently of the LD law.

We then determined the value of $x$, for each LD law and for each object, that minimised the correlation between $u^\prime$ and $v^\prime$. This was done by manual iteration and was restricted to two significant figures in $x$ both for convenience and to avoid unnecessary precision. These values are given in Table~\ref{tab:x} and show that the best value of $x$ depends on both the object and the LD law, as expected. The results are highly consistent, with the exception of IT~Cas for which significantly different $x$ values are found in some cases. A plausible explanation for this is that IT~Cas is the only object with an eccentric orbit, and the inclusion of $e\sin\omega$ as a fitted parameter has modified the correlations between the LD coefficients. However, an exploratory Monte Carlo simulation with $e\sin\omega$ fixed showed the same result so this supposition was not confirmed.

\begin{table} \centering
\caption{\em \label{tab:rev} Linear Pearson correlation coefficients
between $u^\prime$ and $v^\prime$ in our new LD law parameterisations,
calculated for each of the five objects included in the numerical
experimentation using Monte Carlo simulations.}
\begin{tabular}{lcccc}
{\em Object} & {\em quadratic law} & {\em logarithmic law} & {\em square-root law} & {\em cubic law} \\[3pt]
IT~Cas            &  $-$0.860  &  $+$0.969  &  $-$0.888  &  $-$0.842   \\
WASP-50           &  $-$0.034  &  $-$0.375  &  $-$0.890  &  $-$0.028   \\
HAT-P-7           &  $+$0.206  &  $-$0.022  &  $+$0.704  &  $-$0.222   \\
YZ~Cas            &  $-$0.207  &  $+$0.383  &  $-$0.064  &  $-$0.745   \\
WW~Aur            &  $-$0.363  &  $+$0.374  &  $-$0.711  &  $-$0.688   \\[3pt]
\end{tabular}
\vskip10pt
\end{table}

Given this relatively good consistency in $x$, we chose suitable values for implementation in {\sc jktebop} for general use: 0.3 for the cubic law and 0.6 for the other three laws. For clarity, here are the revised versions of the LD laws we propose:
\begin{equation}
u^\prime_{\rm quad} = u_{\rm quad} + 0.6\,v_{\rm quad}
\end{equation}
\begin{equation}
v^\prime_{\rm quad} = u_{\rm quad} - 0.6\,v_{\rm quad}
\end{equation}
for the quadratic law,
\begin{equation}
u^\prime_{\rm log} = u_{\rm log} + 0.6\,v_{\rm log}
\end{equation}
\begin{equation}
v^\prime_{\rm log} = u_{\rm log} - 0.6\,v_{\rm log}
\end{equation}
for the logarithmic law,
\begin{equation}
u^\prime_{\rm sqrt} = u_{\rm sqrt} + 0.6\,v_{\rm sqrt}
\end{equation}
\begin{equation}
v^\prime_{\rm sqrt} = u_{\rm sqrt} - 0.6\,v_{\rm sqrt}
\end{equation}
for the square-root law, and
\begin{equation}
u^\prime_{\rm cub} = u_{\rm cub} + 0.3\,v_{\rm cub}
\end{equation}
\begin{equation}
v^\prime_{\rm cub} = u_{\rm cub} - 0.3\,v_{\rm cub}
\end{equation}
for the cubic law.

\begin{figure}[t] \centering \includegraphics[width=\textwidth]{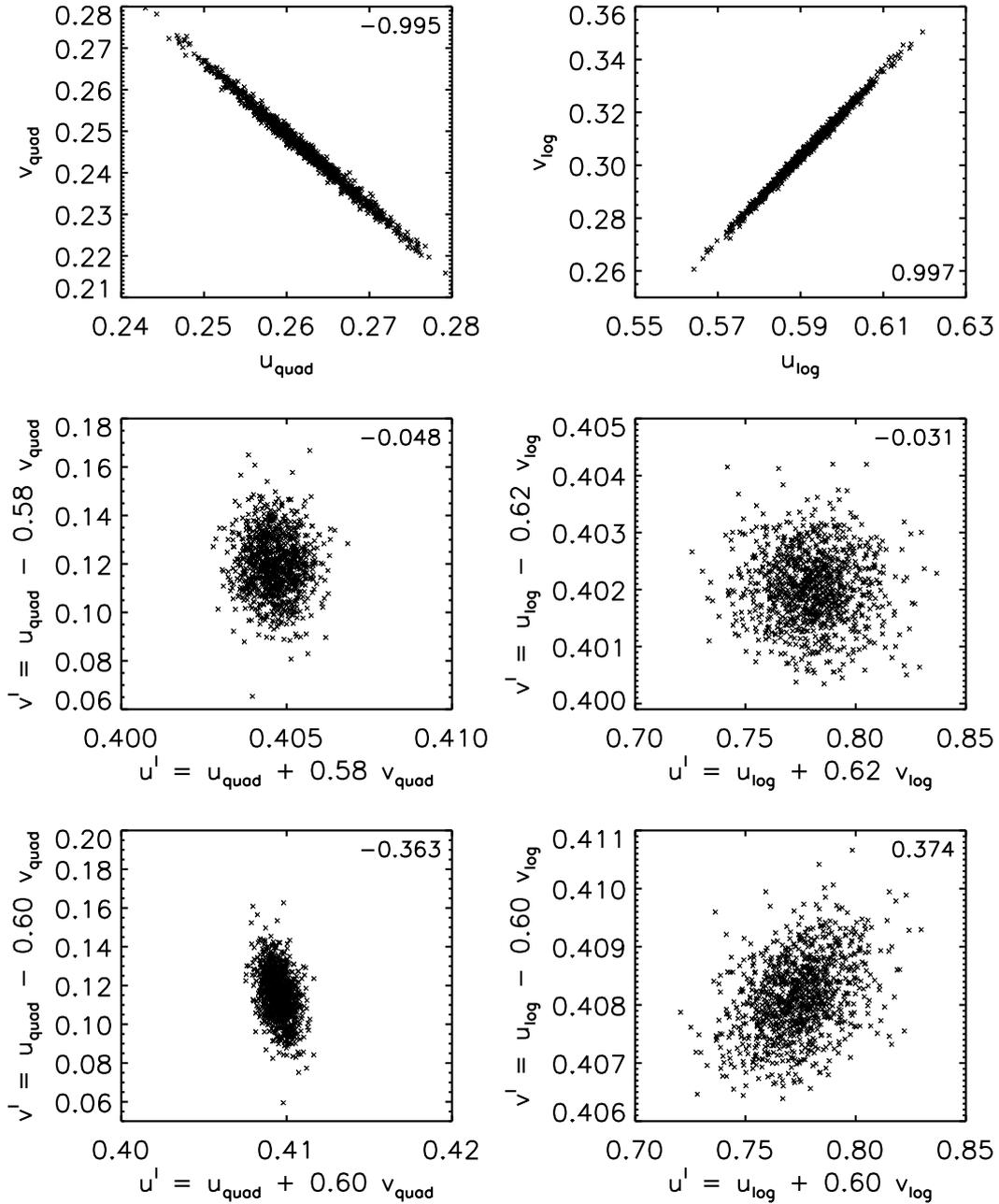} \\
\caption{\label{fig:ww:1} Scatter plots of the LD coefficients for the quadratic and logarithmic
laws obtained from fitting the light curve of WW~Aur and then performing 1000 Monte Carlo
simulations. The correlation coefficient is printed in each panel.} \end{figure}

\begin{figure}[t] \centering \includegraphics[width=\textwidth]{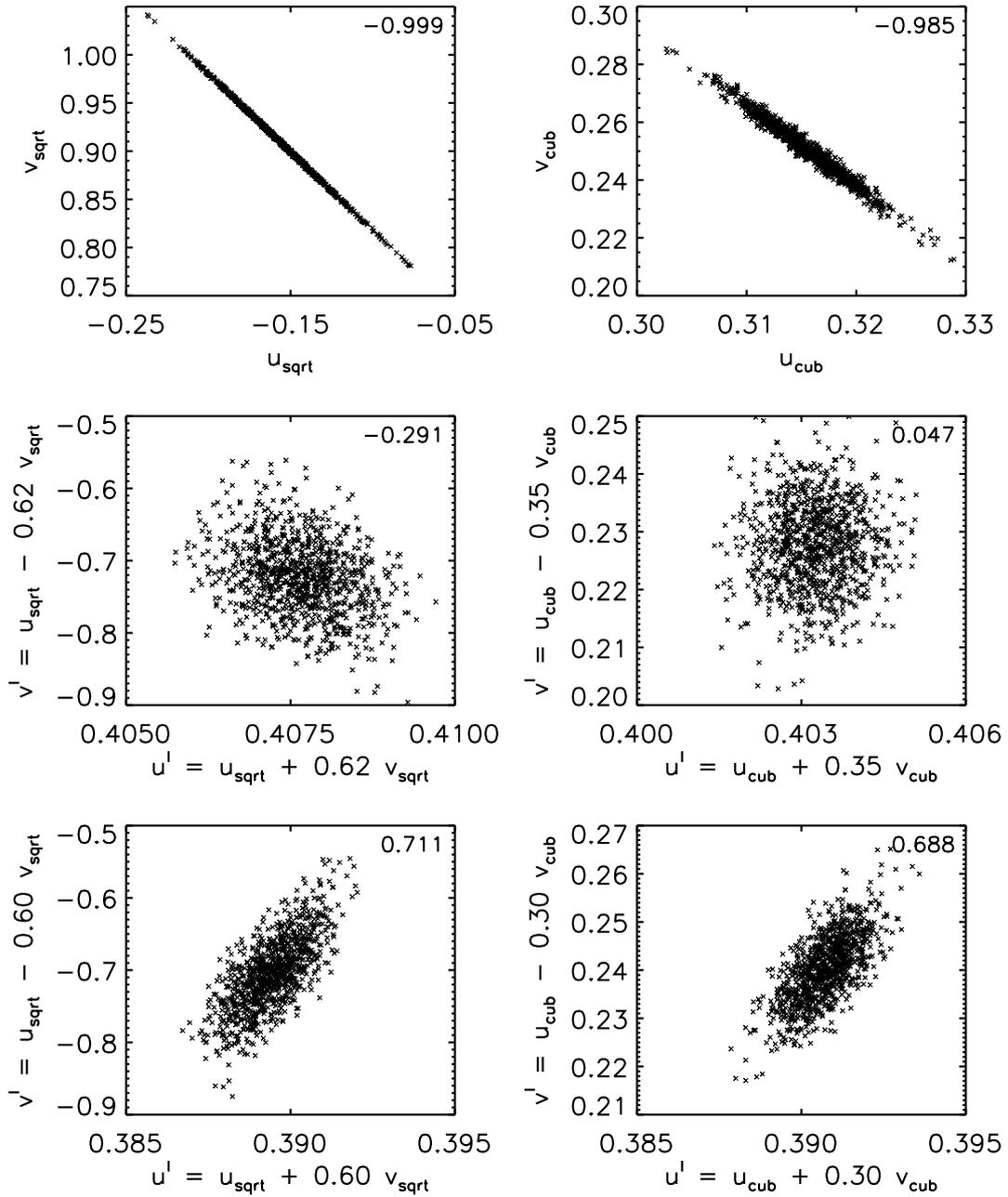} \\
\caption{\label{fig:ww:2} As Fig.~\ref{fig:ww:1} but for the square-root and cubic LD
laws.} \end{figure}

We assessed the correlation between $u^\prime$ and $v^\prime$ for each of these laws and for each of the five objects to gauge the improvement brought by the revised laws. These are given in Table~\ref{tab:rev} and show a clear improvement in all cases. There are nevertheless still some strong correlations, particularly for the logarithmic and square-root laws. We recommend that these laws are not used when attempting to fit both coefficients of a two-parameter LD law. As an example, in Figs. \ref{fig:ww:1} and \ref{fig:ww:2} we show scatter plots of the Monte Carlo simulation output for WW~Aur, for the LD laws in their original form, for the lowest correlation for this object, and for the recommended re-parameterisations.

Several published re-parameterisations of the quadratic LD law \cite{Brown+01apj,Holman+06apj,Pal08mn} were quoted above. We checked these against each of our five objects (allowing for values between 35\degr\ and 40\degr\ for the functional form proposed by P\'al \cite{Pal08mn}) and found that they all yielded significantly stronger correlations than the re-parameterisa- tions proposed in the current work. Finally, we did not attempt to compare the coefficients to theory in order to avoid ``mission creep''.

\section*{Testing the new LD laws}

\begin{table} \centering
\caption{\em \label{tab:wwcompare} Selected results from fitting the TESS light curve
of WW~Aur with one or two LD coefficients fitted, for all possible versions of the one-
and two-parameter laws. $N_{\rm cof}$ is the number of LD coefficients fitted. The
bracketed quantities indicate uncertainties in the final digit of the preceding values}
\begin{tabular}{lccccc}
{\em LD approach}   & $N_{\rm cof}$ & rms (mmag) & $r_{\rm A}$ & $r_{\rm B}$ & $i$ (\degr) \\[3pt]
Linear law                      & 1 & 0.350 & $0.15958~(4)$ & $0.15121~(4)$ & $87.550~(2)$ \\[3pt]
Quadratic law                   & 1 & 0.343 & $0.15973~(4)$ & $0.15148~(4)$ & $87.497~(2)$ \\
Logarithmic law                 & 1 & 0.352 & $0.15957~(4)$ & $0.15118~(4)$ & $87.555~(2)$ \\
Square-root law                 & 1 & 0.341 & $0.15973~(4)$ & $0.15140~(4)$ & $87.508~(2)$ \\
Cubic law                       & 1 & 0.341 & $0.15973~(4)$ & $0.15138~(4)$ & $87.510~(2)$ \\
Power-2 law                     & 1 & 0.341 & $0.15971~(4)$ & $0.15138~(4)$ & $87.512~(2)$ \\[3pt]
Quadratic re-par                & 1 & 0.342 & $0.15972~(4)$ & $0.15146~(4)$ & $87.501~(2)$ \\
Logarithmic re-par              & 1 & 0.647 & $0.16019~(7)$ & $0.15254~(7)$ & $87.300~(4)$ \\
Square-root re-par              & 1 & 0.341 & $0.15973~(4)$ & $0.15140~(4)$ & $87.508~(2)$ \\
Cubic re-par                    & 1 & 0.348 & $0.15960~(4)$ & $0.15124~(4)$ & $87.543~(2)$ \\
Power-2 ($h_1$ and $h_2$)       & 1 & 0.342 & $0.15970~(4)$ & $0.15136~(4)$ & $87.517~(2)$ \\[3pt]
Quadratic law                   & 2 & 0.342 & $0.15969~(4)$ & $0.15141~(4)$ & $87.510~(3)$ \\
Logarithmic law                 & 2 & 0.341 & $0.15972~(4)$ & $0.15141~(4)$ & $87.508~(3)$ \\
Square-root law                 & 2 & 0.341 & $0.15973~(4)$ & $0.15140~(4)$ & $87.508~(3)$ \\
Cubic law                       & 2 & 0.341 & $0.15974~(4)$ & $0.15139~(4)$ & $87.507~(3)$ \\
Power-2 law                     & 2 & 0.341 & $0.15974~(4)$ & $0.15141~(4)$ & $87.507~(3)$ \\[3pt]
Quadratic re-par                & 2 & 0.342 & $0.15969~(4)$ & $0.15141~(5)$ & $87.510~(3)$ \\
Logarithmic re-par              & 2 & 0.341 & $0.15972~(4)$ & $0.15141~(4)$ & $87.508~(3)$ \\
Square-root re-par              & 2 & 0.341 & $0.15973~(4)$ & $0.15140~(5)$ & $87.508~(3)$ \\
Cubic re-par                    & 2 & 0.341 & $0.15973~(4)$ & $0.15140~(4)$ & $87.508~(3)$ \\
Power-2 ($h_1$ and $h_2$)       & 2 & 0.341 & $0.15974~(4)$ & $0.15141~(5)$ & $87.507~(3)$ \\[3pt]
Four-parameter                  & 1 & 0.341 & $0.15970~(4)$ & $0.15136~(4)$ & $87.517~(2)$ \\
Four-parameter                  & 2 & 0.341 & $0.15972~(4)$ & $0.15139~(4)$ & $87.509~(3)$ \\
\end{tabular}
\vskip10pt
\end{table}

Now we had re-parameterisations of the LD laws and implemented them into {\sc jktebop}, we proceeded to test the code and assess the effect of the revised LD laws. To limit the computational load of this work we analysed only one object, WW~Aur, and fitted only the data near eclipse in the first half of the light curve from TESS sector 45. Best fits and 1000 Monte Carlo simulations were performed for the linear LD law, for all two-parameter laws in their original form, for the re-parameterisations presented here, and for the $h_1$ and $h_2$ approach for the power-2 law. Initial or fixed LD coefficients were set to values for the Cousins $R$ passband from Claret \& Hauschildt \cite{ClaretHauschildt03aa}, with the exception of the power-2 law for which we used the TESS passband predictions from Claret \& Southworth \cite{ClaretSouthworth22aa}. We also ran two fits using the four-parameter LD law: one with coefficient $u_2$ fitted and one with $u_2$ and $u_4$ fitted. The values of the fixed coefficients were taken from Claret \cite{Claret17aa}.

We report only the most relevant results from this work: the r.m.s.\ scatter around the best fit, the fractional radii ($r_{\rm A}$ and $r_{\rm B}$), and the orbital inclination ($i$). These are given in Table~\ref{tab:wwcompare} with errorbars assessed using the Monte Carlo simulations. The errorbars are not true uncertainties, as Monte Carlo simulations are only one of the tools typically deployed in our error analyses \cite{Me21obs5}, and are almost certainly too small \cite{Maxted+20mn}. Extensive comparisons between the results from different LD laws can also be found in the supplementary material to our \textit{Homogeneous Studies} papers \cite{Me08mn,Me10mn,Me11mn,Me12mn} for 94 TEPs.

Based on experience, Table~\ref{tab:wwcompare} and the \textit{Homogeneous Studies} supplementary material, we draw the following conclusions. First, the linear LD law is too simplistic and gives slightly different results to those from all other LD laws. It should not be used except for convenience in cases where the data quality is low. Second, the re-parameterised laws give results that are consistent with the original laws. Third, fitting for both LD coefficients yielded comparable results to fitting for one coefficient in the case of WW~Aur, for which the data are of extremely high quality. Fourth, the anomalously poor solution in Table~\ref{tab:wwcompare} for the re-parameterised square-root law suggests that the re-parameterised laws risk giving bad results if only one LD coefficient is fitted and the other coefficient is fixed at a bad value.


\section*{Summary}

A profusion of LD laws have been proposed, many of which have two coefficients. All of these suffer from strong correlations between the two coefficients which hinders the modelling process of observed light curves when both coefficients are fitted parameters. We have proposed a re-parameterisation of the quadratic, logarithmic, square-root and cubic laws and performed numerical simulations to calibrate the re-parameterisations. This was done considering three EBs and two TEPs with a variety of light curve shapes.

We give the following recommendations:
\begin{enumerate}
\item Light curves of low quality can be modelled using either the linear law for simplicity, or one of the two-parameter laws with one or both coefficients fixed.
\item Light curves of medium quality should be modelled using one of the standard two-parameter laws, with one coefficient fitted and one fixed.
\item Light curves of high quality should be modelled including two LD coefficients as fitted parameters. In this case the re-parameterised laws should be used, to avoid the strong correlations found with the standard two-parameter laws. This is particularly important for sampling agorithms such as Markov chain Monte Carlo (MCMC), to avoid long autocorrelation lengths in the Markov chains.
\item If you are unsure whether a light curve is of low, medium or high quality, you should try two or all three options and decide which is best based on the values of and uncertainties in the fitted LD coefficients (and other system parameters).
\item The linear LD law should be avoided when performing any detailed analysis.
\item The quadratic LD law should be avoided as it does not match theoretical LD predictions well \cite{ClaretSouthworth22aa,Neilson+17apj}.
\item The power-2 LD law should be adopted as the default law because it \emph{does} match theoretical LD predictions well \cite{Maxted18aa,ClaretSouthworth22aa,Morello+17aj}.
\item The four two-coefficient LD laws give highly consistent results when treated in the same way, so the choice between them is not important.
\item The best re-parameterisation of a given LD law varies slightly between light curves. If this is an issue, or if you want to avoid parameter correlations as much as possible, you should use principal component analysis (PCA) to orthogonalise the model parameters in the course of obtaining a least-squares solution to a given light curve.
\end{enumerate}

All LD laws and re-parameterisations have been implemented in version 43 of the {\sc jktebop} code, which is freely available for download from the author's website. The choice of which LD function to adopt is left to the user.


\section*{Acknowledgements}

We thank Drs.\ Pierre Maxted and Antonio Claret for comments on a draft of this manuscript.
This paper includes data collected by the \kepler\ and TESS\ missions and obtained from the MAST data archive at the Space Telescope Science Institute (STScI). Funding for the \kepler\ and TESS\ missions is provided by the NASA's Science Mission Directorate. STScI is operated by the Association of Universities for Research in Astronomy, Inc., under NASA contract NAS 5–26555.
 The following resources were used in the course of this work: the NASA Astrophysics Data System; the SIMBAD database operated at CDS, Strasbourg, France; and the ar$\chi$iv scientific paper preprint service operated by Cornell University.


\bibliographystyle{obsmaga}

\end{document}